\magnification=\magstep1
\centerline{\bf Charged Lepton Oscillations and (g-2) Measurements}
\bigskip
\par \noindent
\centerline{A. Widom and Y. N. Srivastava}
\centerline{Physics Department, Northeastern University, Boston MA 02115}
\centerline{and}
\centerline{Physics Department \& INFN, University of Perugia, Perugia Italy}
\vskip 1.5in
\par \noindent
\centerline{\it ABSTRACT}
\bigskip
Traditional analysis of $g=2(1+\kappa )$ experiments for charged leptons 
use a classical spin vector picture. For muons, we here employ a more 
exact Dirac quantum four component spinor theory. Survival probabilities  
(including wave packet effects) are computed. These oscillate with the 
frequency $\Omega =(\kappa eB/Mc)$ as has been assumed in previous muon 
$(g-2)$ experimental analyses; i.e. muon survival probability oscillations 
are already at the root of previous succesful $(g-2)$ measurements. 
Further oscillations should also be observed if mixed neutrino mass 
matrices were to enter into the reactions (e.g. $\pi^+ \to \mu^+ +\nu_\mu$) 
producing muons.
\vfill \eject
\centerline{\bf 1. Introduction}
\medskip

Measurements of the charged lepton anomalous magnetic moments[1-5], 
i.e. the $g$-factor
$$
g=2(1+\kappa ), \eqno(1a)
$$
are of considerable interest. The calculation of $\kappa  $ as a function 
of the coupling strength $\alpha =(e^2/\hbar c)$, 
$$
\kappa  = \big({\alpha \over 2\pi }\big)+...\ , \eqno(1b)
$$
gives one some confidence (when compared with experiment) that 
perturbative quantum electrodynamics (and perhaps some other gauge 
theories) makes sense. New precision measurements of $\kappa $ for the 
muon have been proposed[6] for obtaining additional insights into strong 
and weak interactions[7].  

The analysis of $(g-2)$ measurements often proceeds from classical 
equations[8,9]. For example, a charged lepton is considered to follow
a classical orbit in a magnetic field 
$$
{du^\mu \over d\tau }=\big({e\over Mc}\big)F^{\mu \nu }u_\nu . \eqno(2a)
$$
Even the spin-$1/2$ pseudo-vector is considered to follow a classical 
equation of motion 
$$
{ds^\mu \over d\tau }=\big({(1+\kappa )e\over Mc}\big)F^{\mu \nu }s_\nu 
+\big({\kappa  e u^\mu u_\lambda \over M c^3}\big)
F^{\lambda \nu }s_\nu . \eqno(2b)
$$
Such {\it classical spin} equations can serve only as a rough guide to the 
inherent {\it quantum spin} interference exhibited in laboratory high 
precision measurements of $(g-2)$.

Consider the precision measurements of $(g-2)$ for the muon[2-5]. Muons 
(produced by the decay (say) $\pi^- \to \mu^-+\bar{\nu}_\mu $) are injected 
into a ring with a uniform applied magnetic field ${\bf B}$. The 
experimental quantum survival probability of the muon in the ring 
to decay, (via $\mu^- \to e^- +\bar{\nu}_e+\nu_\mu $ with a detected 
electron energy above a threshold value) has been fit to the 
theoretical functional form 
$$
P_{\mu^- \to e^-+\nu_\mu+\bar{\nu}_e}(t)=e^{-Mc^2 \Gamma t/E}
\big({1+A \cos(\Omega t+\phi)\over 1+Acos\phi}\big). \eqno(3)
$$ 
In Eq.(3), $M$, $E$, and $\Gamma^{-1}$ represent (respectively) the 
mass, energy, and intrinsic lifetime of the muon; $t$ is time in the 
laboratory reference frame, and 
$$
\Omega =\big({\kappa  e B\over Mc}\big) \eqno(4) 
$$ 
is the experimental frequency measured in the muon survival probability 
due to quantum mechanical amplitude interference. The frequency in Eq.(4) 
is central for the experimental determination of $\kappa  $ in Eq.(1). Eq.(3) 
for the muon survival probability (as determined by a detected energetic 
electron) is central for the $(g-2)$ measurement. Eq.(3) represents a 
quantum mechanical charged lepton oscillation which follows from the 
Dirac equation and cannot be fully understood in purely classical terms.
This point will be discussed in detail in the work which follows.
   
Starting from the work of Schwinger[10,11], the magnetic moment anomaly 
parameter $\kappa $ was defined by the manner in which the vacuum 
renormalized lepton mass depended on an applied magnetic field. For 
example, the Dirac-Schwinger equation for a muon moving in a magnetic 
field has the form 
$$
\big(-i\hbar \gamma^\mu d_\mu +{\cal M}c-
i(\hbar \Gamma /2c)\big)\psi(x)=0, \eqno(5a)
$$
where the gauge derivative is defined as   
$$
d_\mu =\partial_\mu -i\big({eA_\mu \over \hbar c}\big),
\ \ F_{\mu \nu}=\partial_\mu A_\nu - \partial_\nu A_\mu, \eqno(5b) 
$$
and the renormalized mass (matrix) has the form 
$$
{\cal M}c^2=Mc^2-\big({\kappa  \hbar e\over 4Mc}\big)
\sigma^{\mu \nu }F_{\mu \nu}. 
\eqno(5c)
$$

Our purpose is to discuss in detail the notion of charged lepton 
oscillations as they have appeared in previous $(g-2)$ measurements 
employing the quantum mechanical Dirac-Schwinger Eqs.(5), rather than 
the classical Eqs.(2). The quantum mechanical superposition of 
amplitudes viewpoint is by far the more fundamental. In Sec.2, the 
Dirac-Schwinger Eqs.(5) will be derived from the usual charged current 
definition of $\kappa $, i.e. 
$$
J^\mu (x)=ec\bar{\psi }(x)\gamma^\mu \psi(x)+
\big({\kappa  \hbar e\over 2M}\big)\partial_\nu 
\big(\bar{\psi }(x)\sigma^{\mu \nu }\psi(x)\big). \eqno(6)
$$   
It will be shown that the second term on the right hand side of 
Eq.(6) gives rise to the second term on the right hand side of 
Eq.(5c). The exact energy spectrum of a charged lepton (with 
$\kappa  \ne 0$) in a uniform magnetic field will be derived in Sec.3 
from the Dirac equation. In Sec.4, the survival amplitude for a muon 
in a magnetic field will be computed from solutions of the Dirac 
equation and the experimental modulation frequency Eqs. (3) and (4) will 
be derived. The general nature of charged muon oscillations 
(including those induced by neutrino oscillations) is discussed in 
the concluding Sec.5. 
\medskip
\centerline{\bf 2. The Dirac-Schwinger Equation}
\medskip

The action for a spin-$1/2$ charged particle may be written as 
$$
S=\int d^4 x \bar{\psi}(i\hbar \gamma^\mu \partial_\mu -Mc)\psi  
+{1\over c^2}\int d^4 x J^\mu A_\mu . \eqno(7)
$$
From Eqs.(6) and (7) it follows that 
$$
S=\int d^4 x \bar{\psi}(i\hbar \gamma^\mu d_\mu -Mc)\psi 
-\big({\kappa  \hbar e\over 2M c^2}\big)\int d^4 x
\bar{\psi }\sigma^{\mu \nu }\psi \partial_\nu A_\mu ,
\eqno(8)
$$
where $d_\mu $ is defined in Eq.(5b) and integration by parts has 
been performed on the second term in Eq.(8). From Eqs.(5c) and (8), 
one may write the action
$$
S=\int d^4 x \bar{\psi}(i\hbar \gamma^\mu d_\mu -{\cal M}c)\psi ,
\eqno(9)
$$
from which the Dirac-Schwinger Eq.(5a) follows. 
However, implicit in the above considerations is the complex mass 
replacement rule 
$$
Mc^2 \to Mc^2-i(\hbar \Gamma /2), \eqno(10)
$$
which will be used throughout this work to describe muon decay 
rates.

For a particle in a uniform magnetic field, the mass matrix in 
Eq.(5c) has two eigenvalues, 
$$
{\cal M}w_\pm ={\cal M}_\pm w_\pm, \eqno(11)
$$
where $w_\pm $ are four component spinors and 
$$
{\cal M}_\pm c^2=Mc^2 \mp \big({\kappa \hbar eB\over 2Mc}\big)
=Mc^2 \mp \big({\hbar \Omega \over 2}\big). \eqno(12)
$$
The mass splitting is thus determined by Eq.(4) as 
$$
\Delta {\cal M}c^2=|{\cal M}_+-{\cal M}_-|c^2=\hbar \Omega . \eqno(13)
$$

In analogy to $K$-meson[12] and $B$-meson physics, there may be a 
temptation to employ a phase shift 
$\theta =-(c^2/\hbar )({\cal M}_+-{\cal M}_-)\tau =\Omega \tau $ in 
terms of the proper time of the classical Eqs.(2). Such temptation 
is ill advised since the experimental phase interference in Eq.(3) 
involves laboratory time $t$ with the phase $\Omega t$ and not the 
``proper time'' $\tau =(Mc^2/E)t$ with phase $\Omega \tau $. This kind 
of error does not arise if one merely solves the Dirac-Schwinger Eqs.(5). 
And this we shall now proceed to do. 
\medskip 
\centerline{\bf 3. Energy Wave Functions}
\medskip 

The Dirac-Schwinger Hamiltonian in the laboratory frame in which there 
is an applied uniform magnetic field is given by
$$
H=c{\bf \alpha \cdot \Pi}+\beta 
\big(Mc^2 -(\kappa \hbar e/2 Mc){\bf \sigma \cdot B}\big), 
\eqno(14)
$$
where 
$$
{\bf \Pi}={\bf p}-(e/c){\bf A},\ \  
[\Pi_i, \Pi_j]=i(\hbar e/c)\epsilon_{ijk}B_k, \eqno(15)
$$
and 
$$
{\bf \alpha}=\gamma_5 {\bf \sigma }. \eqno(16) 
$$
We seek solutions to 
$$
H\psi ({\bf r})=E\psi ({\bf r}). \eqno(17)
$$
With the matrix $\rho_2 $ defined as 
$$
\rho_2 =i\beta \gamma_5,\ \ \rho_2^2=1,\eqno(18a)
$$  
one notes that 
$$
H\rho_2 +\rho_2 H=0. \eqno(18b)
$$
Given a particle solution in Eq.(17) with $E>0$, one may also 
construct anti-particle (negative energy) solutions 
$$
H\rho_2\psi ({\bf r})=-E\rho_2\psi ({\bf r}). \eqno(19)
$$
Here we consider only positive energy ($E>0$) eigenstates.

Note that orbital angular momentum and spin angular momentum about 
the $z$-axis (chosen parallel to the magnetic field) are not 
separately conserved; i.e. with ${\bf L}={\bf r\times p}$ and 
${\bf S}=\hbar {\bf \sigma}/2$, 
$$
[H,S_z]=-[H,L_z]=2ic\gamma_5 ({\bf \Pi \times S})_z. \eqno(20)
$$  
On the other hand the total angular momentum along the magnetic 
field axis is conserved 
$$
J_z=L_z+S_z, \ \ \ [H,J_z]=0. \eqno(21)
$$
Another conserved quantity is $\Pi_z$,
$$
[H,\Pi_z]=0. \eqno(22)
$$
For motions in a plane perpendicular to the magnetic field 
$\Pi_z \psi ({\bf r})=0$, i.e. there is zero motion along the magnetic 
field axis.

Note that for the classical Eqs.(2) of motion in the plane (zero 
motion along the magnetic field axis) we have $(dS_z/dt)_{classical}=0$. 
However, in the quantum mechanical treatment $(dS_z/dt)_{quantum}=
(i/\hbar)[H,S_z]\ne 0$. The classical equations of motion thereby 
contain integrals of motion {\it not} present in the physical quantum 
mechanical treatment. The point is that a particle  with {\it total} 
spin $s=(\hbar /2)$ is {\it not} a classical spinning particle. A classical 
spinning particle has a total spin angular momentum {\it large} on the scale 
of $\hbar $. 

If one employs cylinder coordinates ${\bf r}=(\rho,\phi ,z)$, and the 
Landau gauge for the vector potential in ${\bf B}={\bf \nabla \times A}$,  
$$
A_\rho=0,\ \ A_\phi =(B\rho/2),\ \ A_z=0, \eqno(23)
$$
then for zero motion along the magnetic field axis 
$$
\Pi_z \psi =-i\hbar (\partial \psi /\partial z)=0, \eqno(24)
$$
so that 
$$
\psi =\psi (\rho ,\phi )=u_\downarrow (\rho )e^{im\phi}
+u_\uparrow (\rho )e^{i(m-1)\phi}. \eqno(25)
$$
In Eq.(25), $u_{\downarrow,\uparrow}(\rho )$ are four component spinors 
and conservation of total angular momentum has been invoked; i.e. 
$[H,J_z]=0$ allows us to choose the eigenfunctions in Eq.(17) to also be 
eigenfunctions of $J_z=L_z+S_z$. Thus, the choice  
$$
J_z \psi (\rho ,\phi )=
\hbar \big(m-(1/2)\big)\psi (\rho ,\phi ),\ (m=0,1,2,...)\ , \eqno(26)
$$
determines the coherent superposition of spin up and spin down spinors
in Eq.(25) which enter into high energy experimental measurements of 
$(g-2)$. 

From Eqs.(14), (15), (23), and (25) we find the exact energy spectrum 
$$
E_{\pm ,N}=\sqrt{M_B^2 c^4+2\hbar ecBN\mp(\kappa \hbar eB/M_Bc)  
\sqrt{M_B^2 c^4+2\hbar ecBN}}, \eqno(27a) 
$$
$$
M_B=M\sqrt{1+\big({\kappa \hbar eB\over 2M^2c^3}\big)^2}, \eqno(27b)
$$
where $N=n_\rho +|m|$ and $n_\rho $ is an integer counting nodes in 
the radial wave functions. If there were a momentum component $p_z$ 
parallel to the magnetic field, then Eq.(27) would be modified to  
$$
E_{\pm ,N}(p_z)=
\sqrt{M_B^2 c^4+c^2p_z^2+2\hbar ecBN\mp(\kappa \hbar eB/M_Bc)  
\sqrt{M_B^2 c^4+2\hbar ecBN}}. \eqno(28) 
$$
Eq.(28) describes rigorously the exact energy spectrum of a Dirac 
spin-$1/2$ particle with a $g$-factor defined as $g=2(1+\kappa)$ 
moving in a uniform magnetic field $B=|{\bf B}|$.

For magnetic fields of laboratory size (say less than a few Tesla), 
the inequality $\kappa \hbar eB<<M^2 c^4$ holds by an overwhemingly 
large margin. Thus, {\it very accurately} $M_B$ may be replaced by 
$M$, the liftime of the energy eigenstate to weak muon decays may be 
determined by 
$$
\Gamma_{\pm ,N}=\big({Mc^2\over E_{\pm ,N}}\big)\Gamma , \eqno(29)
$$
and the Bohr transition frequency 
$$
E_{-,N}-E_{+,N}=\hbar \Omega , \eqno(30)
$$
where $\Omega $ is given by Eq.(4) independent of $N$, i.e. 
independent of the energy $E_{\pm ,N}$ itself. In deriving the 
transition rate $\Gamma_{\pm ,N}$ for the weak decay of the muon 
energy eigenstate, Eqs.(10) and (27a) were employed with 
$\hbar \Gamma << Mc^2$ also satisfied again by a very wide margin. 
Within the same high accuracy approximation one may write 
$$
E_{\pm,N+n}-E_{\pm,N}=\big({n\hbar ecB\over E_{\pm,N}}\big),
\ \ \ 1\le |n|<<N. \eqno(31)
$$

Eqs.(29) and (30) constitute the quantum mechanical basis for Eq.(3) 
which lies at the root of the experimental analysis of the muon $(g-2)$ 
measurement. More general statements can be deduced about lepton 
polarization in magnetic fields on the basis of the {\it fundamental 
differences} between the quantum and the classical viewpoints. 
According to the classical Eqs.(2) of motion, it is quite possible 
for a high energy particle with $E>>Mc^2$ and $\kappa \ne 0$ moving in 
a circular orbit in a plane normal to ${\bf B}$ to have the spin 
polarized parallel to the magnetic field. For a quantum Dirac particle 
in a high energy (on the mass scale) eigenstate and with $\kappa \ne 0$, 
the spin polarization is required by the Dirac-Schwinger Hamiltonian to 
be (almost) perpendicular to the magnetic field. We have discussed above 
the reasons for this difference. For the classical equations of motion 
$S_3$ is conserved. For the quantum equations of motion, $S_3$ is 
not conserved. This has an effect on lepton beam polarizations in 
high energy machines, e.g. LEP. We note (in this regard) 
that (at LEP) the electron cyclotron frequency $\Omega_c =(ecB/E)$ 
is certainly not large when compared with the anomaly frequency 
$\Omega =(\kappa eB/mc)$ because even though $\kappa <<1$, the 
energy $E>>mc^2$ is quite high.
\medskip
\centerline{\bf 4. The Muon Survival Amplitude}
\medskip
    
Survival amplitudes are conventionally defined as 
$$
{\cal S}(t)=<\psi (0)|\psi (t)>, \eqno(32) 
$$ 
i.e. the amplitude that a quantum object with a wave function 
$\psi (0)$ at time zero will still be in the same state after a 
time t. We take the scalar product for the muon to be 
$$
{\cal S}(t)=\int d^3r \psi^\dagger ({\bf r},0)\psi({\bf r},t). \eqno(33)
$$ 
For the muon moving in a uniform magnetic field 
$$
\psi({\bf r},t)=\sum_{s=\pm}\sum_N
exp(-\Gamma_{s,N}t/2)exp(-iE_{s,N}t/\hbar)
c_{s,N}\psi_{s,N}({\bf r}), \eqno(34)
$$
so that 
$$
{\cal S}(t)=\sum_{s=\pm}\sum_N |c_{s,N}|^2
exp(-\Gamma_{s,N}t/2)exp(-iE_{s,N}t/\hbar). \eqno(35)
$$

The Survival probability $P(t)=|{\cal S}(t)|^2$ is then 
$$
P(t)=\sum_{s,N,s^\prime N^\prime}|c_{s,N}|^2|c_{s^\prime,N^\prime }|^2
exp(-\gamma_{s,N,s^\prime ,N^\prime}t/2)
cos(\omega_{s,N,s^\prime ,wN^\prime}t), \eqno(36a)
$$
where 
$$
\gamma_{s,N,s^\prime ,N^\prime}=\Gamma_{s,N}+\Gamma_{s^\prime, N^\prime}, 
\ \ \ \hbar \omega_{s,N,s^\prime ,N^\prime}
=E_{s,N}-E_{s^\prime, N^\prime}. \eqno(36b) 
$$
From Eqs.(29), (30), (31), and (36) one finds 
$$
P(t)={1\over 2}\int dW(E) \int dW(E^\prime )
exp(-\gamma_{E,E^\prime }t/2)\big(cos(\omega_{E,E^\prime}t)+
cos(\omega_{E,E^\prime}t+\Omega t)\big), \eqno(37a)
$$
where $dW(E)$ is the probability that the muon has an energy in the 
interval $dE$, 
$$
\gamma_{E,E^\prime }=\Gamma \big({Mc^2\over E}+{Mc^2\over E^\prime}\big), 
\ \ \ \omega_{E,E^\prime }=
\big({eB\over Mc}\big)\big({Mc^2\over E}-{Mc^2\over E^\prime}\big), 
\eqno(37b)
$$
and $\Omega $ is defined in Eq.(4). Eq.(37a) is more simply written as 
$$
P(t)={1\over 2}K(t)(1+cos(\Omega t)), \eqno(38a)
$$
where 
$$
K(t)=\int dW(E) \int dW(E^\prime )
exp(-\gamma_{E,E^\prime }t/2)cos(\omega_{E,E^\prime}t) \eqno(38b)
$$
describes the spreading of the muon wave packet moving around in the 
magnetic field. Such wave packet spreading depends on the precise 
nature of muon energy distribution $dW(E)/dE$. For a Gaussian muon 
energy distribution, with mean energy $\bar{E}=<E>$ and deviations 
from the mean $\delta E=\sqrt{<E^2>-<E>^2}<<\bar{E}$, 
$$
K(t)=exp\big({-\Gamma Mc^2 t\over \bar{E}}\big)
exp\big({-t^2\over 2t_s^2}\big), 
$$
$$
t_s=\big({1 \over \sqrt{2}\delta \omega}\big),
\ \ \big({\delta \omega \over \omega}\big)=
\big({\delta E\over \bar{E}}\big), 
\ \ \omega=\big({eBc\over \bar{E}}\big) . \eqno(38c)
$$

Comparing Eqs.(3) and (4), which have been previously employed[2-5] for 
the analysis of muon $(g-2)$, to our central Eqs.(38) we note some 
similarities and some differences: (i) Our $cos(\Omega t)$ does not 
have an arbitrary phase $\phi $ as in $cos(\Omega t+\phi)$. The reason 
for this is that as theorists we may declare that the initial wave 
function is known at exactly ``time zero''. The experimentalists 
actually have to build the clock and provide the beam (quite different 
from a mere theoretical pronouncement), and ``time zero'' has fluctuated 
a bit in past experiments. This has been reported as a fluctuating (from 
day to day) phase $\phi $. (ii) Eq.(3) has a coefficient $A\le 1$ while 
Eqs.(38) correspond to $A=1$. The reason for this is firstly that we 
have assumed polarization perfectly in the plane perpendicular to 
the magnetic field and nothing in an experiment is perfect. Secondly, 
we have calculated in Eq.(38) the total survival probability while 
Eq.(3) refers to a muon decay ejecting an electron above a high 
energy threshold. The value of $A$ depends at least in part on 
experimental cuts in the reported electron counts. (iii) Eq.(3) contains 
only the intrinsic exponential decay, while our Eqs.(38) also contains 
wave packet spreading effects modeled assuming a Gaussian energy 
distribution. The reason is that wave packet spreading effects have to 
be treated using a quantum mechanical wave (Dirac-Schwinger) equation. 
The spreading of the wave packet is in a time 
$t_s\sim (1/\delta \omega)$. This should not adversely effect the 
accuracy of measured $\kappa $ values for $\delta E/\bar{E}\le 10^{-3}$ 
in laboratory magnetic fields of order $1.5$ Tesla.   
\medskip
\centerline{\bf 5. Conclusions}
\medskip

From the viewpoint of quantum mechanics, the reason for the 
oscillations of any total survival amplitude 
$$
{\cal S}_{tot}(t)=<\Psi |e^{-iHt/\hbar}|\Psi >, \eqno(39)
$$
is in the nature of the energy probability distribution 
$$
dW_{tot}(E)=<\Psi |\delta (E-H)|\Psi >dE , \eqno(40)
$$
i.e. 
$$
{\cal S}_{tot}(t)=\int dW_{tot}(E) e^{-iEt/\hbar }. \eqno(41)
$$

For example, a two peaked Gaussian distribution in energy 
$(\Delta <<\bar{E})$, 
$$
dW_{two\ peak}(E)=(1/2)\sqrt{1/2\pi \Delta^2}
\sum_{s=\pm 1}exp\big(-(E-\bar{E}+s\hbar \omega_o/2)^2/2\Delta^2 \big)
dE ,\eqno(42a)
$$
gives rise to an oscillating in time survival amplitude 
$$
{\cal S}_{two\ peak}(t)=cos(\omega_o t/2)exp(-i\bar{E}t/\hbar ) 
exp(-\Delta^2 t^2/2\hbar^2 ). \eqno(42b)
$$
For another example, if the total energy distribution is a convolution 
of two energy distributions 
$$
{dW_{tot}(E)\over dE}=\int d\epsilon {dW_1(E+\epsilon)\over dE}
{dW_2(\epsilon)\over d\epsilon}, \eqno(43a)
$$
then the survival amplitude is the product of two survival amplitudes 
$$
{\cal S}_{tot}(t)={\cal S}_1(t){\cal S}_2(t). \eqno(43b)
$$
Thus Eq.(38a) represents a convolution of a normal weak muon decay 
with a two peaked (at $\bar{E}\pm (\hbar \Omega/2)$) distribution 
separated by $\Omega =(\kappa eB/Mc)$, i.e. the observed muon 
$(g-2)$ oscillation frequency. 

Consider the notion of neutrino mass mixing[13].
In two previous experiments[14,15], attempts were made to find 
neutrino mass matrix mixing by observing a muon propagating in a 
magnetic field. One tried to measure a double peak in the muon energy 
distribution using a time scale very short compared with $\Omega^{-1}$. 
The point was that the neutrino mass from $\pi^+ \to \mu^+ + \nu_\mu $ 
would induce in the muon energy distribution a double peak from (say) 
two possible masses for the neutrino. These two possible masses would yield 
two possible muon energies, and thus exhibit a double peak in the muon 
energy distribution. Such a double peak induced by neutrino masses did not 
actually appear within the accuracy of the previous experiments. However, 
in those experiments one attempted to measure the muon energy distribution 
{\it directly}. We are here suggesting that a {\it more sensitive} 
probe of energy splitting is of the type already successful in muon
$(g-2)$ experiments; i.e. one should look for the double peak in energy 
via the muon survival probability oscillation in time. The resulting 
extra oscillation should appear in a modified form of Eq.(3) 
$$
P_{\mu^- \to e^-+\nu_\mu+\bar{\nu}_e}(t)\approx e^{-Mc^2 \Gamma t/E}
\big(1-sin^2(2\theta)sin^2(\Omega^\prime t/2)\big)
\big({1+Acos(\Omega t)\over 1+A}\big), \eqno(44a)
$$
where $\theta $ is the two flavor neutrino mixing angle, and
$$
\hbar \Omega^\prime \approx E_\pi \big({\Delta m\over M_\pi}\big)^2. 
\eqno(44b)
$$  
We have assumed that the pion has decayed into a forward moving muon 
(in the laboratory frame) and neutrino with $E_\pi>>M_\pi c^2$. The 
neutrino masses (differing by $\Delta m$) are assumed very small on the 
scale of pion and muon masses. For a freely moving muon (i.e. $\Omega =0$), 
only the neutrino induced oscillations survive as previously discussed[16].

Finally, there are certainly two characteristic frequencies associated 
with $(g-2)$ measurements, i.e. the anomaly frequency 
$\Omega =(\kappa eB/Mc)$ and the cyclotron frequency $\Omega_c =(ecB/E)$. 
The cyclotron frequency $\Omega_c$ has been measured directly from emitted 
radiation. To some extent, the anomaly frequency $\Omega $ should also 
enter into the radiation spectrum, although the authors are not aware of 
any direct measurements of this sort. If there is also neutrino mass mixing, 
then a third frequency $\Omega^\prime $ enters into the problem as well.
\vfill 
\eject
\centerline{\bf References}
\medskip
\par \noindent
1. P. Kusch and H. M. Foley, {\it Phys. Rev.} {\bf 72} (1948) 250. 
\par \noindent
2. G. Charpak, et. al. {\it Il Nuovo Cimento} {\bf 37} (1965) 1241.
\par \noindent
3. J. Bailey, et. al. {\it Il Nuovo Cimento} {\bf 9A} (1972) 369.
\par \noindent
4. F. Combley and E. Picasso, {\it Phys. Rep.} {\bf 14} (1974) 1.
\par \noindent
5. J. Bailey, et. al. {\it Nucl. Phys.} {\bf B150} (1979) 1.
\par \noindent
6. E821 ``Muon g-2 Experiment", Brookhaven National Laboratory USA.
\par \noindent
7. T. Kinoshita and W. J. Marciano in ``Quantum Electrodynamics, 
(Directions in High Energy Physics)'' Vol.7, T. Kinoshita Editor,
World Scientific, (1990) 420.
\par \noindent
8. L. T. Thomas, {\it Phil. Mag.} {\bf 3} (1927) 1. 
\par \noindent
9. V. Bargmann, L. Michel and V. L. Telegdi {\it Phys. Rev. Lett} 
{\bf 2} (1959) 435.
\par \noindent
10. J. Schwinger, {\it Phys. Rev.} {\bf 73} (1948) 416; 
{\it Phys. Rev.} {\bf 75} (1949) 898.    
\par \noindent
11. J. Schwinger, {\it Phys. Rev.} {\bf 82} (1951) 664 (Appendix B). 
\par \noindent
12. T. D. Lee, ``Particle Physics and Introduction to Field Theory'', 
Harwood Academic Publishers, London (1981) 348.
\par \noindent
13. S. M. Bilenky and B. Pontecorvo, {\it Phys. Rep.} {\bf 41} (1978) 
225.
\par \noindent
14. R. Abela, et. al. {\it Phys. Lett.} {\bf B146} (1984) 431.
\par \noindent
15. M. Daum, et. al. {\it Phys. Lett.} {\bf B265} (1991) 425.
\par \noindent
16. Y. N. Srivastava, A. Widom and E. Sassaroli, ``Associated Lepton 
Oscillations'', {\it Proc. Conf. Results and Perspectives in Particle 
Physics}, Le Rencontres de Physique de la Valee d'Aoste, La Thuile, 
Aosta Valley, March 1996 (in press).

\bye